\documentclass[prl,twocolumn,english,showpacs]{revtex4-1}
\usepackage{graphicx}
\usepackage{amssymb}

\begin{document}

\title{Cold heteronuclear atom-ion collisions}

\author{Christoph Zipkes, Stefan Palzer, Lothar Ratschbacher, Carlo Sias$^*$, and Michael K{\"o}hl}

\affiliation{Cavendish Laboratory, University of Cambridge, JJ Thomson Avenue, Cambridge CB3 0HE, United Kingdom}

\begin{abstract}
We study cold heteronuclear atom ion collisions by immersing a trapped single ion into an ultracold atomic cloud. Using ultracold atoms as reaction targets, our measurement is sensitive to elastic collisions with extremely small energy transfer. The observed energy-dependent elastic atom-ion scattering rate deviates significantly from the prediction of Langevin but is in full agreement with the quantum mechanical cross section. Additionally, we characterize inelastic collisions leading to chemical reactions at the single particle level and measure the energy-dependent reaction rate constants. The reaction products are identified by in-trap mass spectrometry, revealing the branching ratio between radiative and non-radiative charge exchange processes.
\end{abstract}


\date{\today}

\maketitle

Cold collisions are characterized by the de-Broglie wavelength of the colliding particles becoming comparable to the length scale of the molecular interactions. Quantum mechanics then dominates the elastic and inelastic scattering phenomena and cross sections. Understanding elastic collisions in this regime is fundamental for achieving and optimizing sympathetic cooling of species which do not allow for direct laser cooling, for example complex molecules. On the other hand, inelastic collisions of translationally cold reaction partners provide a unique opportunity to observe quantum mechanical details of chemical processes which typically are averaged out at higher temperatures \cite{Krems2008}. It is expected that the detailed understanding of cold collisions will eventually pave the way towards coherent chemistry, in which full quantum control of the internal states and reaction pathways will be possible.

Over the past few years, impressive advances have been made in exploring cold collisions of neutral atoms. This has lead, for example, to the controlled creation of weakly bound dimers \cite{Donley2002} and trimers \cite{Kraemer2006} and the production of molecular gases in the rovibrational ground state \cite{Ni2008}. In comparison, cold ion-neutral collisions are much less explored. Ion-neutral reactions offer the outstanding experimental possibility that reactants and reaction products can be observed and manipulated at the single particle level and can be trapped for very long times allowing for precise measurements \cite{Drewsen2004,Staanum2008,Willitsch2008,Staanum2010,Schneider2010}. Collisions of cold (millikelvin) trapped ions with hot ($\sim10^2$\,K) neutral atoms have provided insights into chemical reactions \cite{Staanum2008,Roth2006,Willitsch2008} and play an important role in understanding the molecular composition of the interstellar gas \cite{Smith1992}. Only very recently, the regime in which both collision partners are translationally cold has been accessed by using trapped ions interacting with cold molecular beams \cite{Willitsch2008} or trapped neutral atom samples \cite{Grier2009,Zipkes2010}. Nevertheless, elastic scattering in the quantum regime of many partial waves with its decisive energy dependence of the cross section ($\sigma\propto E^{-1/3}$) was not yet observed. Regarding charge exchange, experimental observations \cite{Grier2009} were well explained by the Langevin model ($\sigma_L \propto E^{-1/2}$) that is based on classical mechanics to describe the mobility of ions in gases \cite{Langevin1905}.

In this manuscript, we explore heteronuclear collision processes between trapped atoms and ions in which both collision partners are translationally cold. We study elastic and reactive collisions using a single ion only in order to better control the kinetic energy than in larger ion crystals. This enables us to investigate the energy dependence of elastic and inelastic scattering processes from 20 to 450 $\mu$eV (equivalent to thermal energies in the range of 0.2 to 5 K). We find the elastic collision rate significantly larger than predicted by Langevin theory but in full agreement with a quantum mechanical calculation. Moreover, we investigate inelastic collisions and measure the energy dependence of the charge exchange cross section and the branching ratio for radiative vs. non-radiative charge exchange. This provides a comprehensive characterization of binary atom-ion collisions in this energy range.


Atom-ion scattering is dominated by the polarization potential, which asymptotically behaves as $V(r)=-\frac{C_4}{2 r^4}$, and a hard-core repulsion at the length scale of the Bohr radius. Here $r$ is the internuclear separation, $C_4=\frac{\alpha q^2}{(4\pi\epsilon_0)^2 }$, $q$ denotes the charge of the ion, $\alpha$ is the dc polarizability of the neutral particle, and $\epsilon_0$ is the vacuum permittivity. Atom-ion scattering was theoretically investigated as early as 1905 when Langevin calculated the drift velocity of ions in a buffer gas. The Langevin model is based on classical mechanics and the predicted cross section is $\sigma_{L}=\pi \sqrt{2 C_4/E}$, leading to an energy independent collision rate constant. Langevin-type collisions occur in close encounters between atom and ion when the impact parameter is below a critical value $b_c=(2 C_4/E)^{1/4}$ \cite{Vogt1954} and exhibit a large momentum transfer between atom and ion. Going beyond the Langevin model, the quantum mechanical description of scattering yields the angular dependence of the differential cross section in more detail. In the energy range under consideration, elastic scattering from the polarization potential involves several partial waves since the s-wave scattering regime of atom-ion collisions is only at sub-microkelvin temperatures \cite{Cote2000,Idziaszek2009}. The semi-classical approximation of the full quantum mechanical elastic scattering cross section for a collision energy $E$ is $\sigma(E)=\pi(1+\pi^2/16)\left(\frac{\mu C_4^2}{\hbar^2  E}\right)^{1/3}$ \cite{Massey1971,Cote2000} ($\mu$ denotes the reduced mass). This includes in particular a large contribution towards forward scattering under very small angles \cite{Zhang2009}, mainly from the centrifugal barrier. The forward scattering peak results from interference of partial waves similar to the bright 'Poisson spot' emerging behind a beam block in wave optics. In these collisions the momentum transfer between the collision partners is extremely small and very difficult to detect. Until now, neither cold ion-neutral collisions nor ion-mobility spectrometry could so far access this regime.

In our experiment we study collisions between ultracold $^{87}$Rb atoms and a single Yb$^+$ ion. We prepare $2.2\times 10^6$ neutral atoms in the $|F=2,m_F=2\rangle$ hyperfine ground state at $T\approx 300$\,nK in a harmonic magnetic trap of characteristic frequencies $(\omega_x,\omega_y,\omega_z)=2 \pi \times (8,28,28)$\,Hz \cite{Palzer2009,Zipkes2010}. We spatially overlap the neutral atoms with a single Yb$^+$ ion trapped and laser cooled in an RF-Paul trap. The ion in the Paul trap experiences a pseudo-potential which is derived from time-averaging a very rapidly oscillating electric quadrupole field ${\bf{E}}(x,y)=\frac{V_{RF} \cos(\Omega t)}{R^2}(x,-y,0)$. $V_{RF}$ denotes the applied voltage, $R$ is the distance of the ion to the electrodes, and $\Omega=2 \pi \times 42.7$\,MHz is the drive frequency. The time-averaged potential provides a harmonic confinement for the ion with characteristic trap frequencies $\omega_\perp=2 \pi \times 150\,$kHz. On top of the time-averaged motion, the ion undergoes very rapid oscillations in position (micromotion) of an amplitude $a_{mm}$ which is proportional to the local electric field strength. The micromotion is a driven oscillation in which the ion's energy is on average conserved because momentum of the ion and the applied force are $\pi/2$ out of phase. Axial confinement of the ion with a characteristic frequency of $\omega_{ax}=2 \pi \times 42$\,kHz is provided by electrostatic fields.

\begin{figure}[htbp]
  \includegraphics[width=0.85\columnwidth,clip=true]{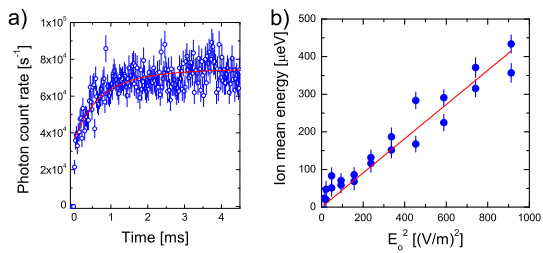}
  \caption{(Color online) a) Time-resolved fluorescence of a single $^{172}$Yb ion. The signal is averaged over 69 repetitions and the error bars are determined by the photon shot noise. The line is a fit \cite{Wesenberg2007} resulting in a mean energy of the ion of $(230\pm9)\,\mu$eV. b) Measured mean ion energy vs. offset electric field used for tuning the excess micromotion. The line is a linear fit to the data.}
  \label{fig1}
\end{figure}


A collision with an ultracold atom changes the instantaneous velocity of the ion and disrupts its coherent micromotion oscillation. As result, energy can be transferred from the ion to the neutral atoms, leading to cooling of the secular motion of the ion \cite{Zipkes2010}, but also energy of up to $\sim \frac{m_{ion}}{2} \Omega^2 a_{mm}^2$ can be transferred from the driving field to the ion's secular motion \cite{Major1968,Devoe2009,Hudson2009}. Depending on the exact parameters, the ion will equilibrate at a mean energy determined by the two processes. We tune the energy of the atom-ion collisions by applying a homogeneous offset electric field $E_o$ transverse to the symmetry axis of the ion trap. This displaces the ion from the geometric center of the Paul trap and introduces excess micromotion of amplitude $a_{mm,ex}= \sqrt{2} q E_o/(m_{ion} \omega_\perp \Omega)$. $m_{ion}$ is the mass of the ion. We center the neutral atomic cloud at the new equilibrium position by applying homogeneous magnetic fields of $\sim 100$\,mG.

We investigate the effect of the excess micromotion $a_{mm,ex}$ on the equilibrated mean energy of the ion after it has interacted with a neutral atom cloud for $t_0 = 8$\,s. During the interaction with the neutral atoms the ion is in its electronic ground state. For the energy measurement we release the neutral atoms from the trap and set the offset electric field to zero. Then, we illuminate the ion with laser light red-detuned by $\Delta=-0.25\,\Gamma$ from the $S_{1/2}\rightarrow P_{1/2}$ transition of Yb$^+$ at 370\,nm, ($\Gamma=2\pi\times 20$\,MHz denotes the linewidth of the atomic transition) and monitor the temporal increase of the fluorescence rate (see Fig. \ref{fig1}a). We fit the data with the solution of the time-dependent optical Bloch equations for the motion of the particle in the pseudopotential and perform a thermal ensemble average \cite{Epstein2007,Wesenberg2007}. In Fig. \ref{fig1}b we display the measurement of the mean energy of the ion vs. the applied offset electric field $E_o$ which exhibits the expected quadratic dependence. We use this energy tuning mechanism to study the energy dependence of the collision cross sections in the next paragraphs. The minimum energy is determined by residual micromotion. We find that the ion equilibrates after less than 100\,ms interaction time to its average kinetic energy.

\begin{figure}[htbp]
\includegraphics[width=.9\columnwidth]{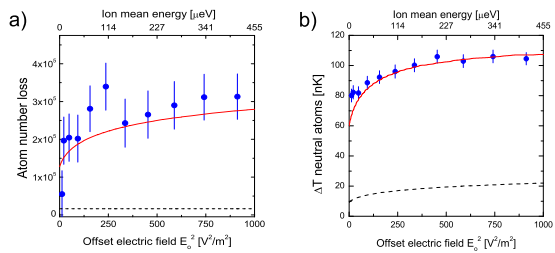}
  \caption{(Color online) a) Atom loss from the magnetic trap vs. mean energy of a single $^{172}$Yb$^+$ ion. The experimental data are averaged over $\sim 70$ realizations each and the standard error is given. The solid line is a numerical simulation based on the full elastic scattering cross section. The dashed line shows the prediction of the Langevin model. b) Temperature increase of the neutral atom cloud vs. mean energy of a single $^{172}$Yb$^+$ ion. The solid line shows the numerical results using the full elastic scattering cross section. The dashed line shows the maximally possible evaporative heating effect. The mean energy of the ion is derived from Fig. \ref{fig1}.}
  \label{fig2}
\end{figure}

The kinematics of a binary collision between an ion of energy $E_{ion}$ and a neutral atom of mass $m_{n}$, which we assume at rest ($E_{n} = 0$), results in a relative energy transfer $\Delta E/E_{ion}=4\gamma/(\gamma+1)^2 \sin^2(\theta/2)$ with $\gamma=m_{ion}/m_{n}$. Collisions in which the scattering angle $\theta$ exceeds $\theta_{cut}=2\arcsin(\sqrt{E_{cut}(\gamma+1)^2/(4 \gamma E_{ion})})$ lead to atom loss from the trap. $E_{cut}$ is the effective trap depth of the atom trap. A neutral atom after a collision will be detected as lost in absorption imaging if it moves at a very large orbit in the trap and does not thermalize with the neutral atom cloud within the time $t_0$. Numerical calculations show that thermalization can be achieved only for energy transfers less than 0.8\,neV. In Fig. \ref{fig2}a we show the measured atom loss as a function of the ion energy. We find that the atom loss exceeds the classical Langevin prediction $\Delta N=n_0\sigma_L v t_0$ by more than an order of magnitude and, moreover, shows a pronounced energy dependence, also not explained by the Langevin model. Here $n_0$ is the local neutral atom density and $v$ is the relative velocity of atom and ion. To model our data, we numerically calculate the quantum mechanical differential elastic scattering cross section $I(\theta,E)$. We calculate the scattering phase shifts in a semi-classical approximation for large values of the angular momentum for a $-C_4/2r^4$ potential and perform a random-phase approximation for small angular momenta for the unknown short-range repulsion \cite{Cote2000}. Using $I(\theta,E)$ we calculate the expected neutral atom losses in a Monte Carlo simulation. The solid line in Fig. \ref{fig2}a shows the predicted loss for large energy transfer ($\Delta E > E_{cut}$).

In the case of small energy transfer ($\theta<\theta_{cut}$) the atom remains in the trap and thermalizes with the remaining atom cloud. The very small amount of transferred energy and momentum ($\theta_{cut}$ is in the mrad-range) make these collisions very challenging to observe. However, using ultracold atoms at nanokelvin temperatures as reaction partners allows us to resolve a relative energy transfer on the order of $\sim 10^{-7}$, inconceivable in previous experiments. The observed temperature increase (Fig. \ref{fig2}b) is explained by two different contributions. Small energy transfers ($\Delta E <E_{cut}$) are dominant. The other effect results from selectively removing atoms from the center of the cloud where their potential energy is below the thermal average, leading to 'evaporative heating'. The dashed line shows the maximally possible contribution from evaporative heating, assuming all observed losses of Fig. \ref{fig2}a being from the center of the trap. The solid line is the predicted temperature increase taking into account both contributions. The simulations in both Fig. \ref{fig2}a and Fig. \ref{fig2}b use the same value of $E_{cut}$ as the only free parameter and we obtain $E_{cut}=0.8$\,neV, in agreement with the experimentally expected value.


Reactive chemical processes in atom-ion collisions, like charge exchange or molecule formation, can only occur when atom and ion approach each other to distances at which the electronic wave functions overlap, which is the case in close encounters. In the regime of many partial waves, the rate constant for chemical processes is thus predicted to be proportional to the Langevin rate constant $\sigma_Lv$ and to be independent of the collision energy \cite{Vogt1954,Cote2000,Makarov2003}. For equal chemical elements strong loss due to near-resonant charge exchange was observed \cite{Grier2009} in agreement with the Langevin estimate.

\begin{figure}[htbp]
\includegraphics[width=.9\columnwidth]{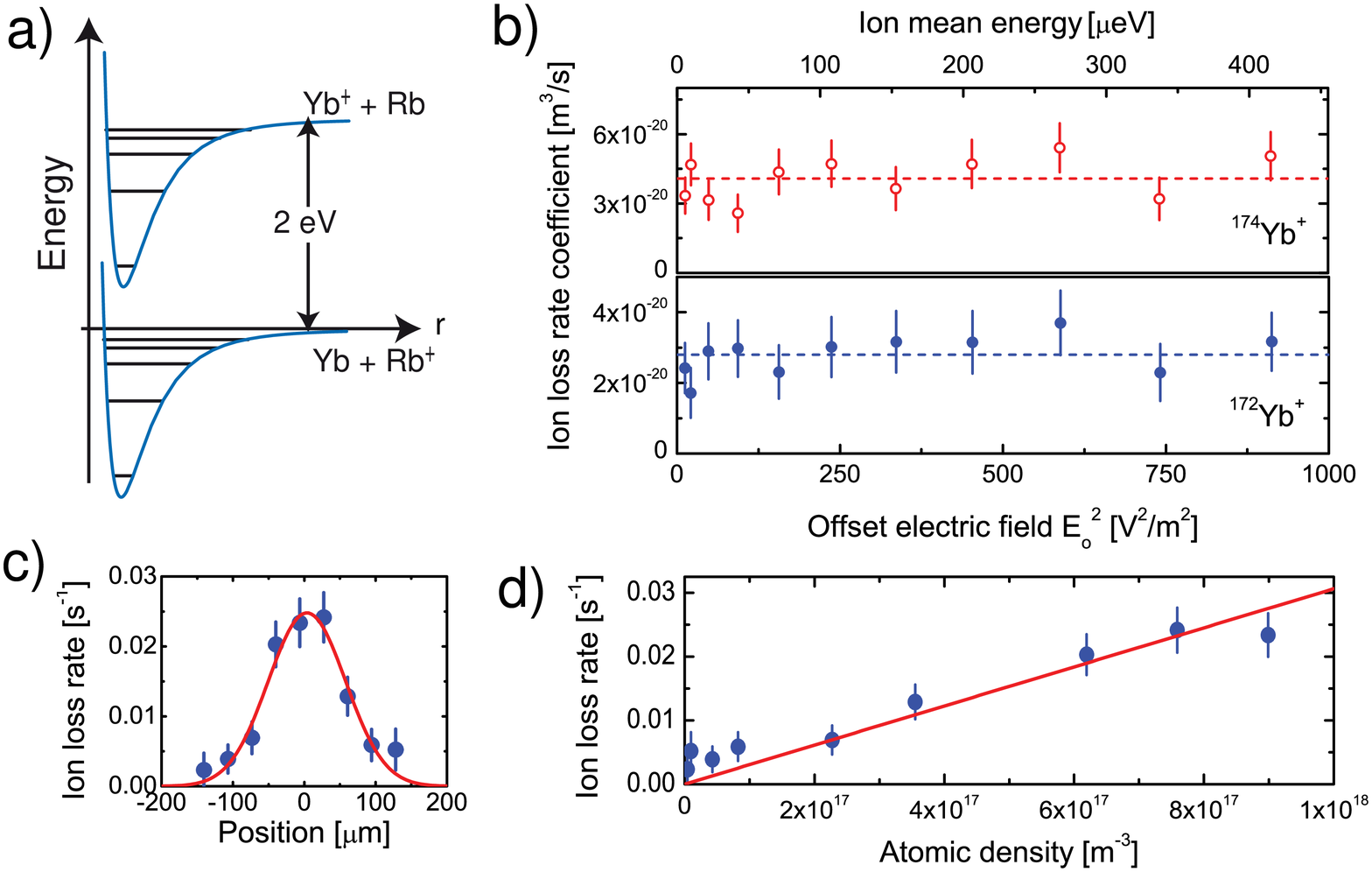}
  \caption{(Color online) {\bf a)} Schematic of the molecular potentials. {\bf b)} Measured ion loss probability as a function of offset electric field. The ion's mean energy is derived from Fig. \ref{fig1}. Each data point is averaged over approximately 100 repetitions of the experiment. The dashed lines show the average of the data. {\bf c)} Position dependence of the ion loss probability for $^{174}$Yb$^+$ as the ion is scanned through the cloud. The red line is the theoretical density profile of the thermal cloud. {\bf d)} The linear dependence of the ion loss rate of the density signals binary collisions leading to charge exchange reactions. Each data point is averaged over approximately 350 repetitions of the experiment.}
  \label{fig3}
\end{figure}

In our case, the Rb ($5s\,^2S_{1/2}$) atom in $|F=2,m_F=2 \rangle$ and the Yb$^+$ ($6s\,^2S_{1/2}$) ion in $|J=1/2,m_J=\pm 1/2\rangle$ collide in the excited A$^1\Sigma^+$ singlet or the a$^3\Sigma^+$ triplet state of the (RbYb)$^+$ molecular potential (see Fig. \ref{fig3}a). The electronic ground state $X^1\Sigma^+$, which asymptotically corresponds to Rb$^+$ ($4p^6\,^1S_0$) and neutral Yb($6s^2\,^1S_0$), lies 2.08\,eV below. The most striking experimental observation of a chemical process is the disappearance of the Yb$^+$ fluorescence which we detect after the interaction with the neutral atoms. Using the same ion energy tuning mechanism as described above, we have investigated the energy dependence of the reaction rate constant. For two different isotopes, $^{172}$Yb and $^{174}$Yb, we find the rate constant to be independent of the collision energy, as predicted \cite{Langevin1905,Vogt1954} (see Fig. \ref{fig3}b). The average rate constants of $R_{172}=(2.8\pm0.3)\times10^{-20}$\,m$^3$/s and $R_{174}=(4.0\pm0.3)\times10^{-20}$\,m$^3$/s are five orders of magnitude smaller than in the homonuclear case \cite{Grier2009} and of the same order of magnitude as predicted for the similarly asymmetric system (NaCa)$^+$ \cite{Makarov2003}. There is a systematic error between the measurements for the two isotopes of 15$\%$ due to uncertainty of the density determination of the thermal cloud.

Additionally, we have measured the density dependence of the inelastic atom-ion collisions by positioning the ion at different locations inside the neutral atom cloud (Fig. \ref{fig3}c). We achieve this without increasing the ion's micromotion by applying a homogeneous magnetic bias field to displace the position of the neutral atomic cloud. Monitoring the ion loss rate as a function of the position, we find that the inelastic atom-ion collisions scale linearly with the local atomic density, indicating that the charge exchange reactions are binary atom-ion collisions (see Fig. \ref{fig3}d). This also demonstrates the principal capability of the single ion as a local probe for density measurements in a neutral atom cloud \cite{Kollath2007}.

\begin{figure}[htbp]
\includegraphics[width=0.9\columnwidth]{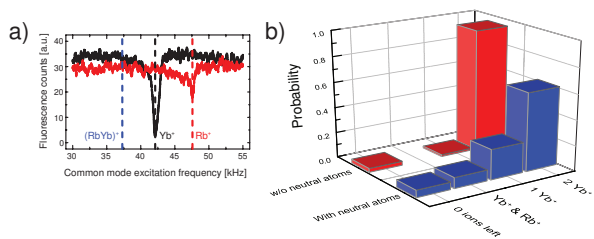}
  \caption{(Color online) a) Mass spectrometry signal together with dashed lines indicating the expected frequencies. The black trace corresponds to two $^{174}$Yb$^+$ ions. For one Rb$^+$ ion and one Yb$^+$ ion (red curve) the collective mode frequency is higher. No signals at the resonance for Yb$^+$ and (RbYb)$^+$ were obtained. b) Observed distribution of the reaction products for two Yb$^+$ ions after 8\,s interaction time with the neutral Rb atoms (blue bars, 486 events). The red bars (543 events) show a comparative measurement without neutral atoms.}
  \label{fig4}
\end{figure}

Charge exchange can occur by emission of a photon (radiative charge exchange) or as a nonadiabatic transition between molecular levels. For low temperatures, radiative charge exchange has been predicted to be the dominating process for (NaCa)$^+$ \cite{Makarov2003}. In order to investigate the reaction products of the charge exchange process we perform mass spectrometry. To this end, we load two Yb$^+$ ions into the ion trap and overlap them with the neutral cloud. In the cases in which only one of the two ions undergoes a reaction, the other ion serves to identify the reaction product by measuring a common vibrational mode in the ion trap \cite{Drewsen2004,Willitsch2008}. We excite the axial secular motion of Yb$^+$ by applying intensity modulated light at 370\,nm at an angle of 60$^\circ$ relative to the symmetry axis of the Paul trap and monitor the fluorescence. The frequency of the intensity modulation is swept linearly from 30\,kHz to 55\,kHz in 2.5\,s. When the intensity modulation coincides with a collective resonance of the secular motion, the ions are heated and the fluorescence reduces (see Fig. \ref{fig4}a). The collective mode for a Rb$^+$ and an Yb$^+$ ion in the trap is $13\%$ above the bare Yb$^+$ mode at 42\,kHz, whereas the mode of (RbYb)$^+$ and Yb$^+$ would be $12\%$ lower in frequency. Fig. \ref{fig4}b shows the histogram for the observed charge exchange processes. In the cases in which one Yb$^+$ ion is lost, we find approximately 30$\%$ probability for the production of cold Rb$^+$ and $70\%$ for a complete loss. If the reaction products take up 2\,eV as kinetic energy in a non-radiative charge exchange they would be lost due to finite depth ($\sim 150$\,meV) of our ion trap. We have not observed the formation of a charged molecule in this process.

In conclusion, we have provided a comprehensive survey of cold heteronuclear atom-ion collisions. We have observed that Langevin theory is insufficient to describe cold atom-ion collisions but the full quantum mechanical cross section is required to describe our data. Moreover, we have observed charge exchange reactions between a single ion and ultracold atoms and analyzed their products. Our results provide an excellent starting point for future experiments targeting the full quantum control of chemical reactions at the single particle level. Photoassociation or magnetic field induced Feshbach resonances \cite{Idziaszek2009} could be used to create single trapped cold molecules in specific rovibrational states.

We thank D. Meschede, G. Shlyapnikov , and V. Vuletic for discussions and {EPSRC} (EP/F016379/1, EP/H005676/1), {ERC} (Grant number 240335), and the Herchel Smith Fund ({C.S.}) for support.

\end{document}